\documentclass[RNAAS]{aastex63}
\usepackage{acronym}
\usepackage{url}

\begin{document}

\title{Fastcc: fast colour corrections for broadband radio telescope data}
\correspondingauthor{M.~W.~Peel}
\email{email@mikepeel.net}
\author[0000-0003-3412-2586]{Mike W. Peel}
\affiliation{Instituto de Astrof\'{i}sica de Canarias, 38200 La Laguna, Tenerife, Spain}
\affiliation{Departamento de Astrof\'{i}sica, Universidad de La Laguna, 38206 La Laguna, Tenerife, Spain}

\author[0000-0001-5479-0034]{Ricardo Genova-Santos}
\affiliation{Instituto de Astrof\'{i}sica de Canarias, 38200 La Laguna, Tenerife, Spain}
\affiliation{Departamento de Astrof\'{i}sica, Universidad de La Laguna, 38206 La Laguna, Tenerife, Spain}

\author[0000-0002-0045-442X]{C.~Dickinson}
\affiliation{Jodrell Bank Centre for Astrophysics, Department of Physics and Astronomy, The University of Manchester, Oxford Road, Manchester M13 9PL, UK}

\author[0000-0003-2514-9592]{J.~P.~Leahy}
\affiliation{Jodrell Bank Centre for Astrophysics, Department of Physics and Astronomy, The University of Manchester, Oxford Road, Manchester M13 9PL, UK}

\author[0000-0002-6439-5385]{Carlos L\'opez-Caraballo}
\affiliation{Instituto de Astrof\'{i}sica de Canarias, 38200 La Laguna, Tenerife, Spain}
\affiliation{Departamento de Astrof\'{i}sica, Universidad de La Laguna, 38206 La Laguna, Tenerife, Spain}

\author[0000-0002-6805-9100]{M. Fern\'andez-Torreiro}
\affiliation{Instituto de Astrof\'{i}sica de Canarias, 38200 La Laguna, Tenerife, Spain}
\affiliation{Departamento de Astrof\'{i}sica, Universidad de La Laguna, 38206 La Laguna, Tenerife, Spain}

\author[0000-0001-5289-3021]{J.~A. Rubi\~{n}o-Mart\'{\i}n}
\affiliation{Instituto de Astrof\'{i}sica de Canarias, 38200 La Laguna, Tenerife, Spain}
\affiliation{Departamento de Astrof\'{i}sica, Universidad de La Laguna, 38206 La Laguna, Tenerife, Spain}

\author[0000-0002-9941-2077]{Locke D. Spencer}
\affiliation{Institute for Space Imaging Science, University of Lethbridge, Lethbridge, Alberta, T1K 3M4, Canada}

\received{}
\accepted{}
\published{}

\begin{abstract}
    Broadband receiver data need colour corrections applying to correct for the different source spectra across their wide bandwidths. The full integration over a receiver bandpass may be computationally expensive and redundant when repeated many times. Colour corrections can be applied, however, using a simple quadratic fit based on the full integration instead. Here we describe \texttt{fastcc} and \texttt{interpcc}, quick Python and IDL codes that return, respectively, colour correction coefficients for different power-law spectral indices and modified black bodies for various Cosmic Microwave Background related experiments. The codes are publicly available, and can be easily extended to support additional telescopes.
    
    \vspace{0.5em}
    \noindent\textit{Unified Astronomy Thesaurus concepts:} \href{http://astrothesaurus.org/uat/259}{CMBR detectors (259)}
    \href{http://astrothesaurus.org/uat/1355}{Radio receivers (1355)}
    \href{http://astrothesaurus.org/uat/1858}{Astronomy data analysis (1858)}
    \href{http://astrothesaurus.org/uat/1234}{Photometry (1234)}
\end{abstract}

\section{Background}

Cosmic Microwave Background (CMB) observations require telescopes using high sensitivity receivers with wide bandwidths that are a large fraction of the observing frequency. For example, the 28.4\,GHz channel of the {\it Planck} satellite was sensitive to photons at frequencies between $\sim24-34$\,GHz (see Fig.~\ref{fig:thefig}), a fractional bandwidth (i.e., $\Delta\nu/\nu$) of $\sim$35\,\%. With such high fractional bandwidths, the measured flux density of an astronomical source depends on the convolution of the observing spectral bandpass with the source spectrum. For example, a bandpass that is most sensitive at low frequencies will result in an increase in the measured flux density of a steep spectrum radio source compared to a flatter spectrum source. Colour corrections avoid these effects by correcting the flux density to that which a monochromatic receiver operating at a reference frequency $\nu_0$ would have measured.

The colour correction thus depends on the spectral index of the source, and if the value for this changes (e.g., while fitting a model with multiple spectral components, or when adding new observations), then the colour correction also changes. For this reason, raw flux densities are published without colour corrections, but colour corrections are applied for any scientific analysis and for the inclusion of data in data visualization and subsequent analysis.

The \texttt{fastcc} and \texttt{interpcc} codes provide fast colour corrections for CMB datasets, particularly in MCMC fitting techniques.


\section{Formalism}
\begin{figure}[tb]
    \centering
    \includegraphics[width=0.49\textwidth]{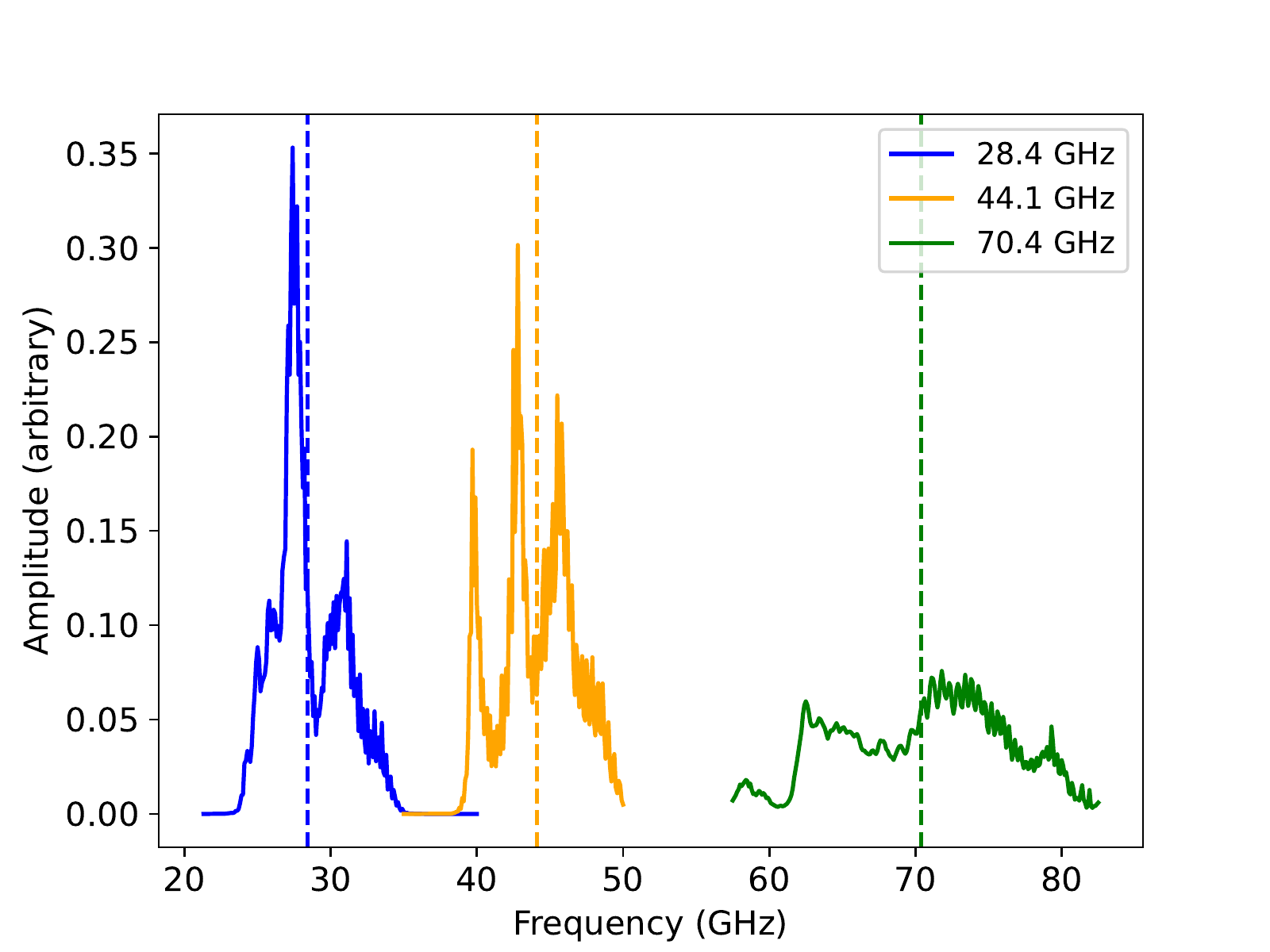}
    \includegraphics[width=0.49\textwidth]{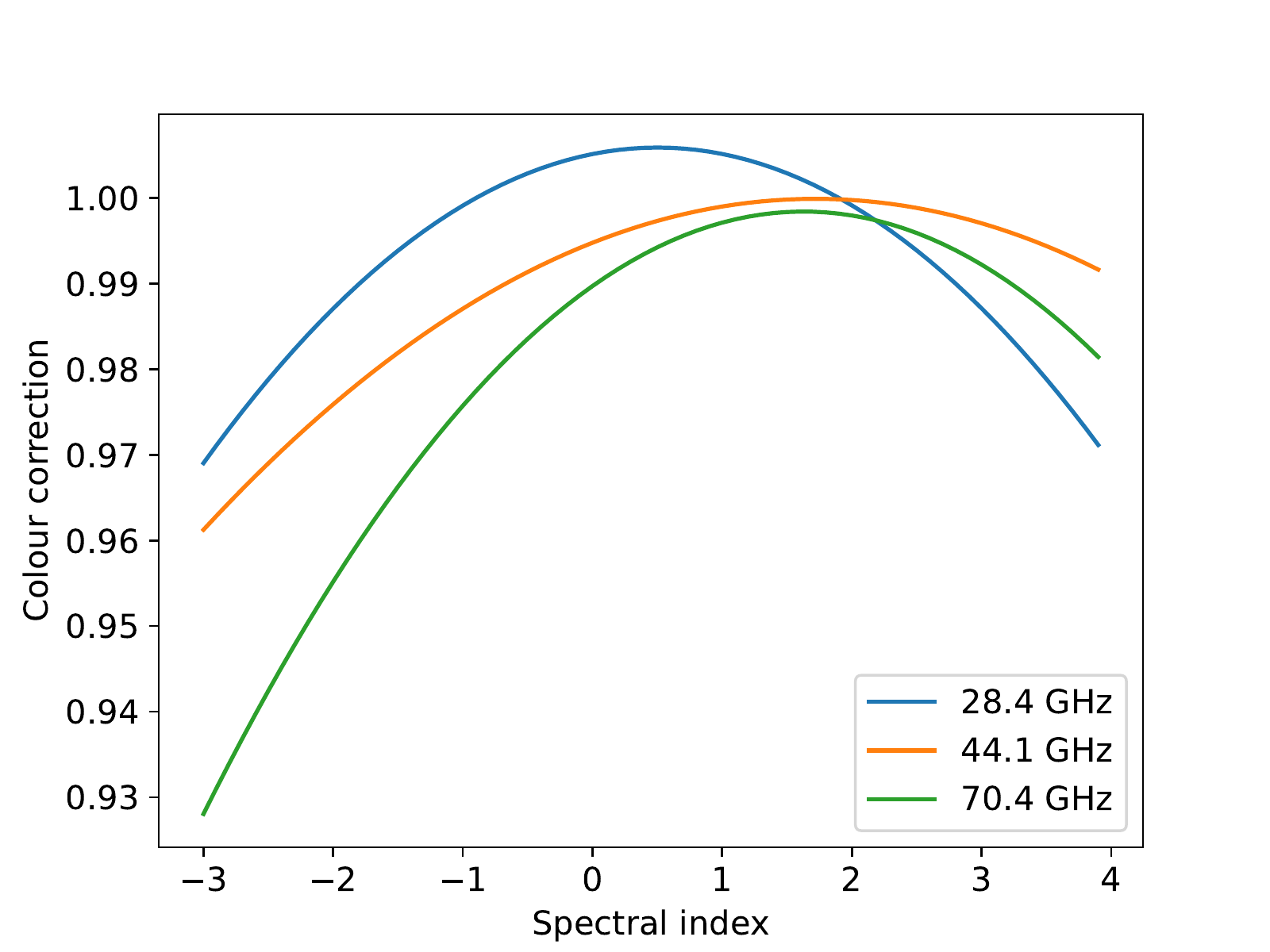}
    \caption{\textit{Left}:
    example bandpasses
    from Planck LFI, showing how these can be complex structures that extend significantly beyond the reference frequency for broadband detectors (dashed lines). \textit{Right}: corresponding colour correction fits for
    various spectral indexes.
    }
    \label{fig:thefig}
\end{figure}
We follow and expand on the formalism from \citet{2016A&A...594A...2P}.
When the source spectrum is described by a power-law, $S \propto \nu^\alpha$, the colour correction is the ratio of integrals over frequency $\nu$, with the numerator integrating over the bandpass $g(\nu)$ (if measured in temperature units)  and the spectrum the data is calibrated to, $\delta$, and the denominator integrating over the combination of the bandpass and the source spectrum with spectral index $\alpha$, as

\begin{equation}
C(\nu_0,\alpha) = \frac{\int g(\nu)(\nu/\nu_0)^{\delta-2}~d\nu}{\int g(\nu)(\nu/\nu_0)^{\alpha-2} ~d\nu},
\end{equation}
where $\nu_0$ is the reference frequency. If the instrument bandpass has been measured in intensity units, $\tau(\nu)$, this becomes:
\begin{equation}
C(\nu_0,\alpha) = \frac{\int \tau(\nu)(\nu/\nu_0)^{\delta}~d\nu}{\int \tau(\nu)(\nu/\nu_0)^{\alpha} ~d\nu}.
\end{equation}
For instruments calibrated in CMB thermodynamic units, this becomes:
\begin{equation}
C(\nu_0,\alpha) = \frac{\int g(\nu)\eta_{\Delta T}  (\nu)/\eta_{\Delta T}(\nu_0)~d\nu}{\int g(\nu) \left(\nu/\nu_0\right)^{\alpha-2}~d\nu},
\label{eq:c_pl}
\end{equation}
where $\eta_{\Delta T}(\nu) = x^2 e^x/(e^x-1)^2$, with  $x=h\nu/(k_{\rm B} T_{\rm CMB})$ to adjust for the CMB spectrum in Rayleigh–Jeans temperature units, with Planck constant $h$, Boltzmann constant $k_{\rm B}$, and CMB temperature $T_{\rm CMB}=2.7255$\,K. However, the ratio of $\eta_{\Delta T}$ is usually small,
only becoming important when $\nu_0$ is far from the centre of the bandpass.

The resulting colour corrections are very smooth (see Fig.\,\ref{fig:thefig}, right), and
can be fitted with a quadratic function:
\begin{equation}
C(\nu_0,\alpha) = c_0 + c_1\alpha + c_2\alpha^2 .
\end{equation}
The source flux density is then corrected as:
\begin{equation}
\label{eq:fluxCor}
    S_\mathrm{corr}(\nu_0) = S_\mathrm{orig}(\nu_0) C(\nu_0,\alpha) = S_\mathrm{orig}(\nu_0) \times \left( c_0 + c_1\alpha + c_2\alpha^2 \right) .
\end{equation}


The colour corrections for thermal dust
spectra can be calculated using the same equations as for power laws, but changing the source spectrum for a 
grey body,
with the dust temperature $T_\mathrm{dust}$ and spectral index $\beta_\mathrm{dust}$ as input parameters. For equations, see the \texttt{fastcc} explanatory supplement. A quadratic fit becomes increasingly different from the colour corrections at extreme values. As such, we also include \texttt{interpcc}, which uses interpolation over a set of cached colour correction values.

There are two equivalent approaches to colour correction: either change the flux density of the source at a fixed reference frequency, or change the reference frequency for the observations. When characterising multiple sources within a survey, the analysis is normally simpler if all sources are at the same frequency with corrected flux density. Alternatively, effective frequencies $\nu_\mathrm{eff}$, for which $C(\nu_\mathrm{eff},\alpha)=1$, can be calculated from the colour corrections  (in thermodynamic units: convert to intensity at $\nu_\mathrm{eff}$) via:
\begin{equation}
\nu_\mathrm{eff} = \nu_0 C(\nu_0,\alpha)^{1/(\alpha-2)}.
\end{equation}

\section{Code}
The code takes input $\alpha$ for \texttt{fastcc}, or $T_\mathrm{dust}$, and $\beta_\mathrm{dust}$ for \texttt{interpcc}, and returns the factor that the measured flux density must be \textbf{multiplied} to obtain the corrected flux density. We include both Python and IDL versions of \texttt{fastcc}, and \texttt{interpcc} in Python.
Test/example scripts are also included for both \texttt{fastcc} and \texttt{interpcc}. The code is available at \url{https://github.com/mpeel/fastcc} and \url{https://doi.org/10.5281/zenodo.7376510}, along with a more detailed explanatory supplement. An IDL version for \textit{Planck} LFI only was released in the Planck Legacy Archive.\footnote{\url{http://pla.esac.esa.int/pla/}} While no dependencies are required for \texttt{fastcc}, \texttt{interpcc} and the colour correction calculations use \texttt{Astropy} \citep{astropy:2018},
\texttt{Numpy} \citep{harris2020array}, and \texttt{scipy} \citep{2020SciPy-NMeth}.

The code now includes colour corrections for \textit{Planck} LFI \citep{2014A&A...571A...5P,2014A&A...571A...9P,2016A&A...594A...2P} and HFI \citep{2014A&A...571A...9P}, WMAP \citep{2013ApJS..208...20B}, IRAS \citep{1988iras....1.....B},  DIRBE,\footnote{\url{https://lambda.gsfc.nasa.gov/product/cobe/dirbe_ancil_sr_get.cfm}} QUIJOTE MFI 
(Rubi{\~n}o-Mart{\'{\i}}n et al., MNRAS accepted, Genova-Santos et al., in prep.), and C-BASS (\citealp{2014MNRAS.438.2426K}, Taylor et al., in prep.). Additional experiments can be added on request and provision of bandpasses.

\bibliography{bibliography}
\bibliographystyle{aasjournal}

\end{document}